# 3D characterization of ultrasonic melt processing on the microstructural refinement of Al-Cu alloys by synchrotron X-ray tomography

Yuliang Zhao [a, b, c*], Dongfu Song [b, d], Bo Lin [e], Chun Zhang [c, f], Donghai Zheng [a], Zhi Wang [b], Weiwen Zhang [b*]

[a] School of Mechanical Engineering, Dongguan University of Technology, Dongguan, 523808, China

[b] National Engineering Research Centre of Near-net-shape Forming for Metallic Materials, South China University of Technology, Guangzhou, 510641, China

[c] School of Engineering & Computer Science, University of Hull, East Yorkshire, HU6 7RX, UK

[d] Guangdong Institute of Materials and Processing, Guangzhou, 510650, China

[e] School of Mechanical Engineering, Guizhou University, Guiyang, 550025, China

[f] School of Aeronautics, Northwestern Polytechnical University, Xi'an, 710072, China

Corresponding author: zhaoyl@dgut.edu.cn (Y. Zhao); mewzhang@scut.edu.cn (W. Zhang)

**Abstract**

The effect of ultrasonic melting processing (USP) on three-dimensional (3D) architecture of intermetallic phases and pores in two multicomponent cast Al–5.0Cu–0.6Mn–0.5(1.0)Fe alloys is characterized using conventional microscopy and synchrotron X-ray microtomography. The two alloys are found to contain intermetallic phases such as α-$Al_{15}(FeMn)_3Cu_2$, β-$Al_7Cu_2Fe$, $Al_3(FeMn)$, $Al_6(FeMn)$, and $Al_2Cu$ that have complex networked morphology in 3D. The application of USP in alloys can obtained refined and equiaxed microstructures. The grain size of 0.5Fe and 1.0Fe alloys is greatly decreased from 16.9 μm, 15.8 μm without USP to 13.3 μm, 12.2 μm with USP, respectively. The results show that USP significantly reduce the volume fraction, grain size, interconnectivity, and equivalent diameter of the intermetallic phases in both alloys. The volume fraction of pores in both alloys is reduced due to the USP degassing effect. The refinement mechanism of USP induced fragmentation of primary and secondary dendrites via acoustic bubbles and acoustic streaming flow were discussed.

1. **Introduction**

Aluminium (Al) alloys as one kind of most widely used metallic alloys in the automobile and aerospace field owing to the high specific strength, high corrosion resistance, and excellent recyclability [1]. Iron is the common impurity in the commercial Al alloys, which make a great challenge to the aluminium recycling [2]. Recycled Al-Cu casting alloys are widely used in the transportation industry by reduced vehicle weight, fuel consumption and greenhouse gas emission [3]. With the accumulation of iron could be resulting in the formation of hard and brittle Fe-rich intermetallic phases (named Fe-rich phases hereafter), especially the plate-like β-$Al_5FeSi$ and β-$Al_7Cu_2Fe$, which have the detrimental effect on the castability and mechanical properties [4]. Thus, the authors [5, 6] have been carried out to modify the morphology of Fe-rich intermetallic phase from plate-like to less-harmful Chinese script through addition the neutralize element, *i.e.*, Mn and Si. However, additions of neutralizing elements increase the volume fraction of Fe-rich intermetallic phases and decrease the toughness of the alloys. The physical external field [7-9], such as ultrasonic field and magnetic field, is a kind of sustainable and economical method for improving the quality and soundness of casting ingots. The application of ultrasonic melt processing (USP) in Al melting during solidification has the advantage of grain refinement and reduce porosity and improve structural homogeneity [10]. It is generally recognized that the mechanisms of cavitation-enhanced nucleation and cavitation-induced fragmentation contribution to the grain refinement of alloys [8, 9]. Therefore, several attempts have been made to alter the morphology and distribution of intermetallic phases, such as Fe-rich phases in Al alloys, through the application of USP [11, 12]. Understanding the morphology and distribution of Fe-rich intermetallic phases can provide the better way to evaluate the properties of the alloys. Presently, mostly researches to reveal the Fe-rich intermetallic phases of samples is characterised by 2-dimensions (2D) images using optical or electron microscopy. However, it is different to fully reveal the complicated 3-dimensional (3D) morphologies and spatial structure of the Fe-rich phases.

Combined transmission electron microscopy (TEM) and focused ion beam (FIB) tomography methods have been used to study the 3D morphology of Fe-rich phases in Al-Si alloys and found that the Chinese-script α-$Al_{15}(FeMn)_3Si_2$ phase has a highly concentrated branched network structure in 3D morphology [20]. The application of serial sectioning method in 3D rendered the structures of Fe-rich phases and found that USP can effectively refined the Fe-rich phases [21]. However, FIB is normally used for sectioning sub-micrometre features and serial sectioning is often very time-consuming. Thanks to the advantage of high brilliance third-generation synchrotron X-rays sources with improving spatial and temporal resolution [13], which provide powerful technique to study the microscale particles, such as Fe-phases in present study. Because of the non-destructive characteristics when X-rays penetrate deeply into materials, synchrotron radiation X-ray computed tomography (SRXCT) has proved to be the unique and powerful method for 3D characterization of microstructure of a wide variety of Al alloys [14]. Recently progress reported on the 3D characterisation of Fe-rich phases using

SRXCT is mainly focused in Al-Si alloys [15-19] and found the Fe-rich phases had a plate-like morphology. Recently, the highly interconnected 3D morphology Fe-rich intermetallic phases were reported in Al-Cu alloys [22], while they only focused in low Fe content (0.1% wt.). Plate-like $Al_3(MnFe)$ and β-Fe and Chinese script $Al_6(MnFe)$ and α-Fe may appear in Al-Cu alloys [23], which are quite different from those in Al-Si alloys. The 3D morphology of α-Al dendrites arms of AA6082 alloy containing $Y_2O_3$ composites appear much thinner, more branched and curved by applying USP using in-situ SRXCT methods [24]. The SRXCT results indicated that the 3D distribution of copper sulfate particles in solid polymer has been changed by USP [25]. To date, few detailed investigations of 3D morphology of Fe-rich phases in Al-Cu alloys and the effect of USP on them. Furthermore, the mechanisms of USP treatment causes 3D morphology changes of Fe-rich phases are not fully understood.

In the present study, we using OM, SEM and SRXCT technique to comparatively study the 3D structures and morphologies of Fe-rich phases, $Al_2Cu$ and pores formed in Al–5.0Cu–0.6Mn–0.5(1.0)Fe alloys. For the first time, we clearly demonstrated USP change the 3D morphologies and distributions of the Fe-rich phases, $Al_2Cu$ and pores. From the experimental results, the mechanisms of grain refinement and reduced porosity induced by USP were established.

## 2. Experiment and methods

2.1 Materials

The design composition of two alloys were Al-5.0Cu-0.6Mn-0.5Fe (0.5Fe) and Al-5.0Cu-0.6Mn-1.0Fe (1.0Fe, in wt.%). The actual chemical composition of alloys was determined by optical emission spectroscopy, the results is listed in Table 1. These alloys were prepared using the commercial pure Al ingot (99.9%), Al-20Cu master alloy, Al-10Mn master alloy and Al-10Fe master alloy with the correct charge weight. The liquidus temperature of the two alloys was determined to be approximately 657 and 656 °C, respectively [26]. To prepare these two alloys, the raw materials were firstly melted at 780 °C in a clay-graphite crucible in an electric furnace, then the melting was degassed by 0.5% $C_2Cl_6$. After removed the slag, holding the melting temperature at 710 °C for 3 min, then cast into the steel permanent mould preheated to 200 °C.

Table 1 Chemical composition of design alloys

| Alloys | Cu | Mn | Fe | Al |
|---|---|---|---|---|
| Al-0.5Mn-0.6Mn-0.5Fe (0.5Fe) | 5.19 | 0.61 | 0.55 | Bal. |
| Al-0.5Mn-0.6Mn-1.0Fe (1.0Fe) | 5.43 | 0.62 | 1.00 | Bal. |

2.2 Ultrasonic melt processing procedure

The ultrasonic system consists of a 1-kW ultrasonic generator, a 20-KHz magnetostrictive transducer and a titanium ultrasonic sonotrode. After the melting poured into the mould, the ultrasonic generator was started work. Ultrasonic processing was performed at 900 W power with the frequency of ~18 kHz and the amplitude of 20 μm. The titanium horn was preheated to 200 °C by using resistant heating coil, this step is to minimize the chill effect of cold horn. According to determined experiment data, the whole solidification time for the two alloys was ~42.5 s with an average cooling rate of ~2.5 K/s [26]. So, the USP time of 30 s means that UPS process the melt from above liquidus temperature to solid temperature. The detailed information about the USP can be found in our previous work [27]. The cylindrical ingots with size of Φ 65 mm × 68 mm were obtained.

The ultrasonic intensity ($I$) can be quantified by the following equation [9]:

$$I = \frac{1}{2}\rho c (2\pi fA)^2 \tag{1}$$

where $\rho$ is the density of the Al melt (2.43 g cm$^{-3}$ [28]), and $c$, $f$, and $A$ are the velocity (4.7 × 10$^3$ m s$^{-1}$ [29]), frequency (19 kHz), and amplitude (20 μm) of the ultrasound wave inside the Al-5Cu melt, respectively. Thus, the ultrasonic intensity was equal to 3255 W cm$^{-2}$.

2.3 Microstructural observation

Cylindrical samples (~ Φ 10 mm × 20 mm) were extracted from the position near the tip of ultrasound horn. Routine 2D microstructure characterisation was made using a LEICA/DMI5000M optical microscope. Volume fraction statistics for different phases was performed by Image J software [30] and each condition was calculated using 30 images and average out. Samples for polarized optical observation were examined in LEICA optical microscope after anodizing with a 4% $HBF_4$ solution for about 30 s at 20 V. Crystal orientation and grain size map were obtained using automated EBSD on a Nova SEM 430 equipped with HKL Channel 5 EBSD acquisition system. Before the sample was performed EBSD, they were polished by Ion Beam Milling System Leica EM TIC 3X machine at 5 kV for 2 h and 3 kV for 0.5 h.

2.4 Synchrotron X-ray tomography

Samples for tomography scans were machined into cylindrical shape with a size of Φ 2 mm × 5 mm. These images were obtained using polychromatic hard X-rays at the TOMCAT beamline X02DA of the Swiss Light Source (Paul Scherrer Institut, Switzerland). The imaging system consists of a 100 μm LuAG: Ce scintillator coupled to a white-beam compatible microscope with a 6.8 × magnification. X-rays generated by the synchrotron penetrate into the sample and converted into visible light by the scintillator. The detector was GigaFRoST and the effective pixel size was 1.62 μm. The angular projection step was set to 2000 projections over 180° leading to the exposure time of 7.0 ms for 4 samples. A series of raw images were obtained, and these images were converted to cross-sectional slices GridRec algorithm. 500 projection (16-bit) were reconstructed into $500^3$ voxels with a voxel size of $(1.62\ micron)^3$. The Image J [30] software used to adjust the contrast between Fe-rich phases and $Al_2Cu$ phase. The 3D segmentation and volume rendering were performed using Avizo Lite v9.0.1 (VSG,

France). The 3D bilateral filter was applied to the raw images in order to increase the contrast and reduce noise. The detailed image processing procedure is shown in Fig. 1.

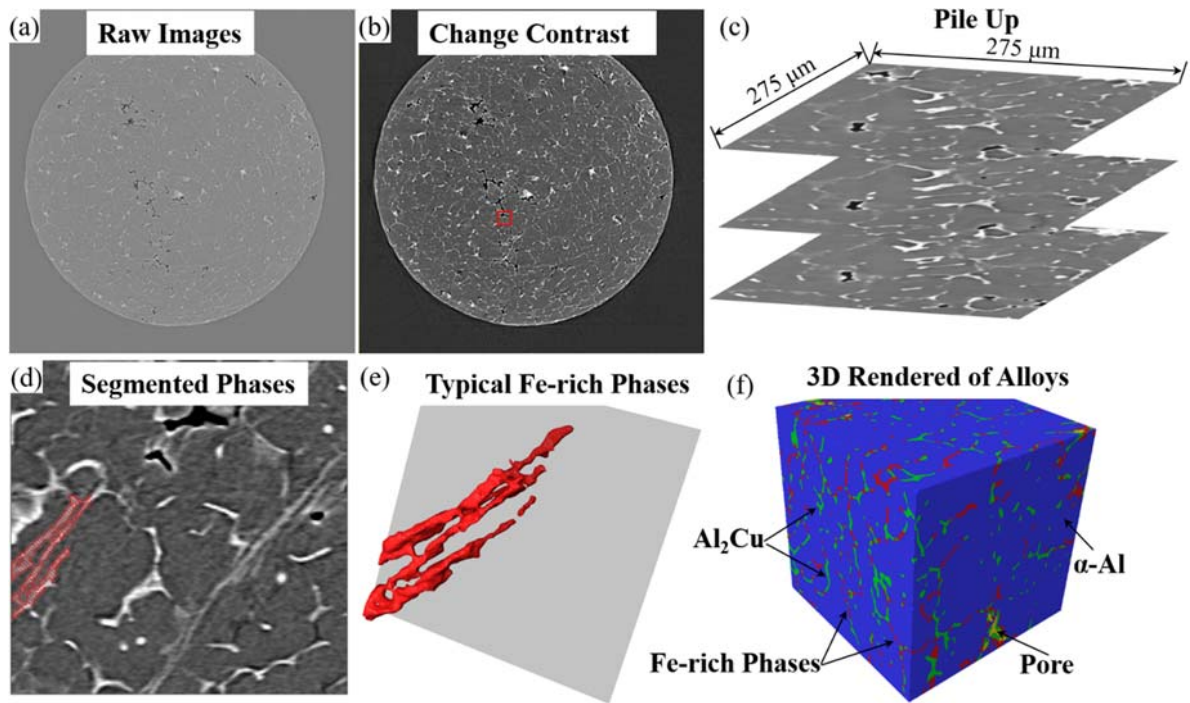

Fig. 1 Detailed image processing procedure: (a) a typical raw image from the tomography scan in 1.0Fe alloy; (b) images processed using a 3D bilateral filter; (c) pile up the images in the Avizo software; (d) segmented the phases through different threshold values; (e) a 3D rendered typical Fe-rich phases image; (f) 3D rendered different phases and pores in 1.0Fe alloy.

We used the equivalent diameter $D_{eq}$, interconnectivity $I$ and specific surface area $SSA$ to quantify the size and shape of the Fe phases. The equivalent diameters $Deq$ is defined as average equivalent diameters of the different particles of the considered phase and the volume V represents the whole volume of considered phase [31]

$$D_{eq} = \sqrt[3]{\frac{6V}{\pi}} \qquad (2)$$

The interconnectivity $I$ is the ratio between the largest 3D individual volume ($V_{larg}$) and the total volume ($V$) [32]

$$I = \frac{V_{larg}}{V} \tag{3}$$

The other important parameter specific surface area *SSA* is defined as surface area *A* per unit volume *V* [33]:

$$SSA = \frac{A}{V} \tag{4}$$

The mean curvature *H* can be characterized using the two principal radii of curvature, $R_1$ and $R_2$, as follows [34]:

$$H = 0.5 * \left(\frac{1}{R_1} + \frac{1}{R_2}\right) \tag{5}$$

The skeletonization function insert in the Avizo® software using the Centerline Tree technique to simplify the 3D structures of the phases into 1-voxel-thick skeletons with connected nodes [34]. Thus, the node length between two nodes can be calculated, and the 3D morphology of the phases can be quantitatively analysed.

## 3. Results and Discussion

**3.1 Effect of USP on the 2D microstructure of the alloys**

Fig. 2 shows the OM microstructure of as-cast 0.5Fe and 1.0Fe alloys without and with USP. The alloys without USP contain dendritic microstructures, Fe-rich phases and $Al_2Cu$ (Fig. 2a and c), whereas equiaxed and homogenous microstructures are observed in the alloys with USP (Fig. 2b and d). The volume fraction of pores observed in the alloys with USP was less than those of alloys without USP due to the ultrasonic degassing effect. The type of intermetallic phases was identified based on chemical composition obtained from EDX and literature [22, 23]. The eutectic intermetallic phases including Fe-rich phases and $Al_2Cu$. 0.5Fe alloy contain α-$Al_{15}(FeMn)_3Cu_2$ and β-$Al_7Cu_2Fe$, while 1.0Fe alloy presents $Al_3(FeMn)$ and $Al_6(FeMn)$. It was seen that the USP reduced the size of intermetallic phases and pores (Fig. 2).

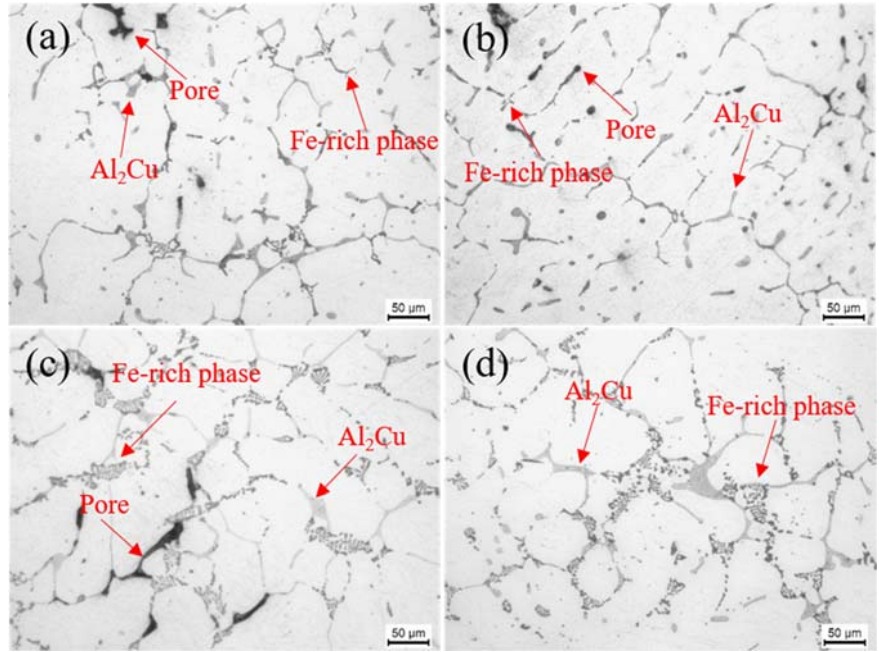

Fig. 2 Optical micrography (OM) of the microstructure: (a) 0.5Fe alloy without USP; (b) 0.5Fe alloy with USP; (c) 1.0Fe alloy without USP; (d) 1.0Fe alloy with USP.

From the polarised light micrographs of the Al-Cu-Mn-Fe alloys without and with USP, it can be seen that the microstructure is mainly composed of primary α-Al dendrites and interdendrite intermetallic phases (Fig. 3). It is evident see in Fig. 3a and c that the large columnar grains and pores were observed in microstructure without USP. However, after applying USP, the microstructure in consists of equiaxed grains and significant grain refinement in are observed in Fig. 3b and d. Furthermore, the secondary dendrite arm spacing (SDAS) of α-Al dendrites processed by USP is obviously smaller than those without USP. For the purpose of investigating the effect of USP on the grain size of alloys, the EBSD orientation maps were presented in Fig. 4. The EBSD show similar results that USP have a significant grain refinement effect on the microconstituents of the alloys. It was proposed that the refinement was mainly due to the cavitation bubbles and acoustic streaming flow induced the fragmentation of α-Al dendrites and intermetallic phases [8, 9].

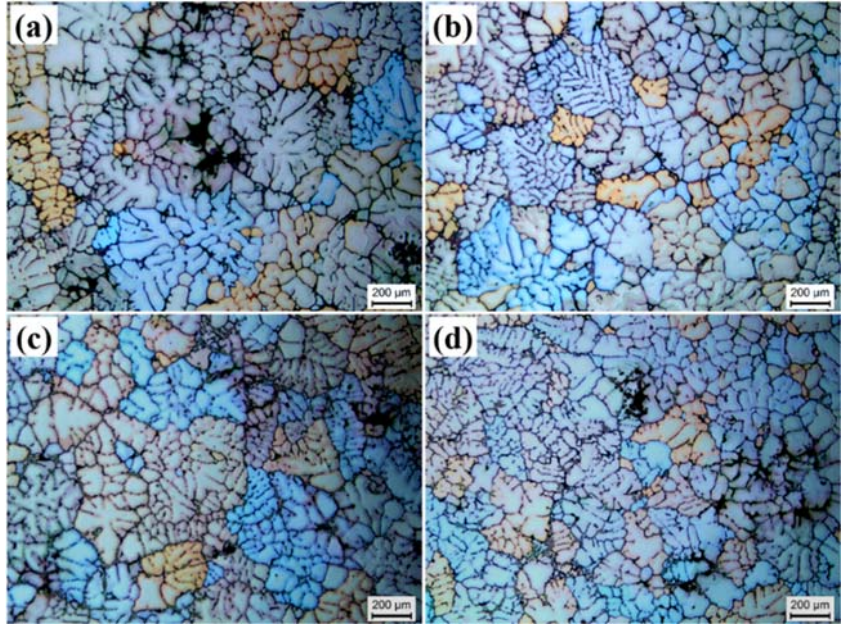

Fig. 3 Polarised light micrographs of the alloys: (a) 0.5Fe alloy without USP; (b) 0.5Fe alloy with USP; (c) 1.0Fe alloy without USP; (a) 1.0Fe alloy with USP.

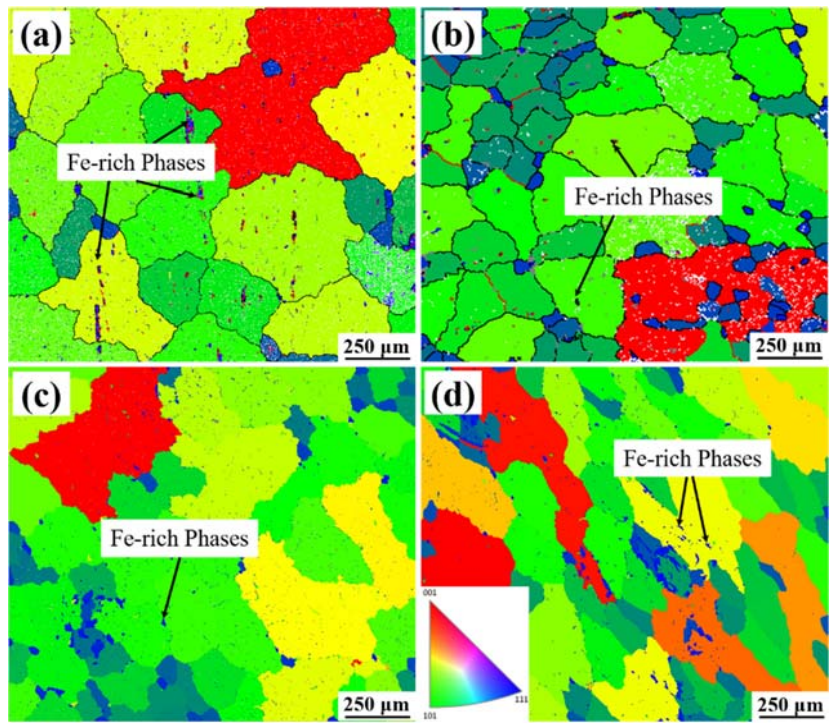

Fig. 4 EBSD shows the grain sizes of the alloys: (a) 0.5Fe alloy without USP; (b) 0.5Fe alloy with USP; (c) 1.0Fe alloy without USP; (a) 1.0Fe alloy with USP.

The secondary dendrite arm spacing (SDAS) and grain sizes of the Al-5Cu-0.6Mn-0.5(1.0)Fe alloys obtained from polarised light micrographs and EBSD results were quantitatively measured and the results are illustrated in Fig. 5. It is obvious that the SDAS and grain size were greatly decreased by USP. Note that SDAS decreased with the application of USP, changing from 16.9 μm, 15.8 μm in the case of 0.5Fe and 1.0Fe alloys without USP to 13.3 μm, 12.2 μm, respectively, after applying USP. With regard to grain size, the value of 416 μm and 336 μm in the case of 0.5Fe and 1.0Fe alloys without USP decreased to 187 μm and 204 μm after applying USP, respectively. Thus, this further confirming that grain refinement was induced by the applying of USP.

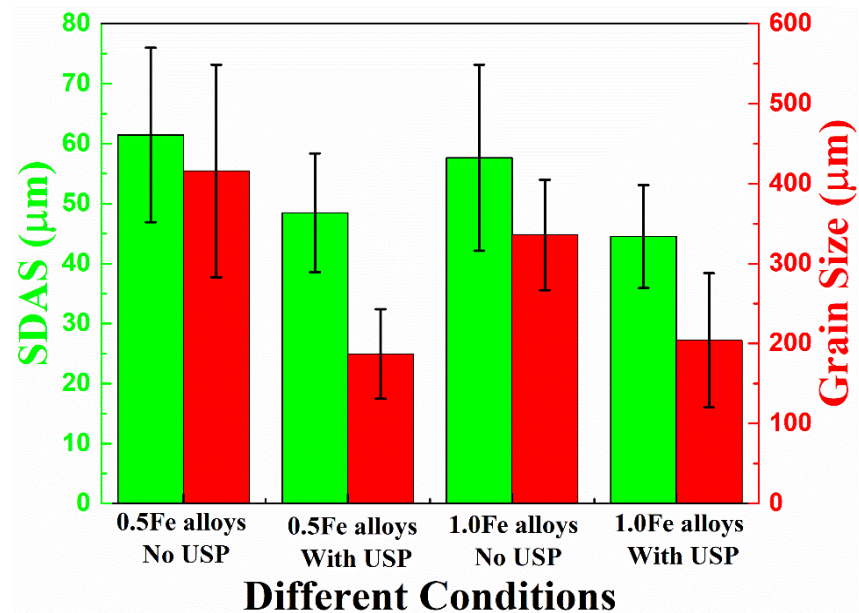

Fig. 5 Variation of secondary dendrite arm spacing (SDAS) and grain size as alloys with and without USP.

Fig. 6 shows the SEM images of deeply-etched Fe-rich phases, *i.e.*, β-$Al_7Cu_2Fe$, α-$Al_{15}(FeMn)_3Cu_2$, $Al_3(FeMn)$ and $Al_6(FeMn)$, in the alloys without and with USP. It can be seen that the 3D morphology of Fe-rich phases without USP is obviously larger than those of with USP. This similar result can be found in the Refs [35, 36]. As shown in Fig. 6a, the 3D morphology plate-like β-$Al_7Cu_2Fe$ shows "hollow" characteristics, owing to the coupled eutectic reaction with α-Al: L → α-Al + α-$Al_{15}(FeMn)_3Cu_2$ + β-$Al_7Cu_2Fe$ + $Al_2Cu$ [6, 23] at 542-537 ºC. After USP, 3D morphology of β-$Al_7Cu_2Fe$ become much thinner (Fig. 6e). It is

worthy of note that α-$Al_{15}$(FeMn)$_3$$Cu_2$ is the typical interconnected Chinese-script intermetallic phase formed through peritectic reaction: L + $Al_6$(FeMn) → α-Al + α-$Al_{15}$(FeMn)$_3$$Cu_2$ at 589-597 °C [23], which are located in the inter-dendritic region and isolated by the (Al) dendrites (marked as blue dotted line in Fig. 6b and f). It is clear that the branch of α-$Al_{15}$(FeMn)$_3$$Cu_2$ become much more compact after application of USP, this is due to the acoustic cavitation and streaming flow [8, 9] break the primary α-Al dendrite and α-$Al_{15}$(FeMn)$_3$$Cu_2$ formed in the inter-dendritic region at the later stage of solidification. The 3D morphology of the $Al_3$(FeMn) phases were often recognized as "needle-shaped phases" (Fig. 6c), while it is more or less like a long rod-like structure with a few small connected branches from the vertical direction (Fig. 6g). $Al_3$(FeMn) is formed through the eutectic reaction: L → α-Al + $Al_3$(FeMn). Fig. 6d and h reveals that $Al_6$(FeMn) in 3D is more or less complex plate-like morphology. This complex morphology is because $Al_3$(FeMn) transforms to $Al_6$(FeMn) through the peritectic reaction: L + $Al_3$(FeMn) → α-Al + $Al_6$(FeMn) at 589-597 °C [23]. $Al_6$(FeMn) inherit the 3D morphology characteristics of $Al_3$(FeMn), which create the nucleation site for the $Al_6$(FeMn) phase.

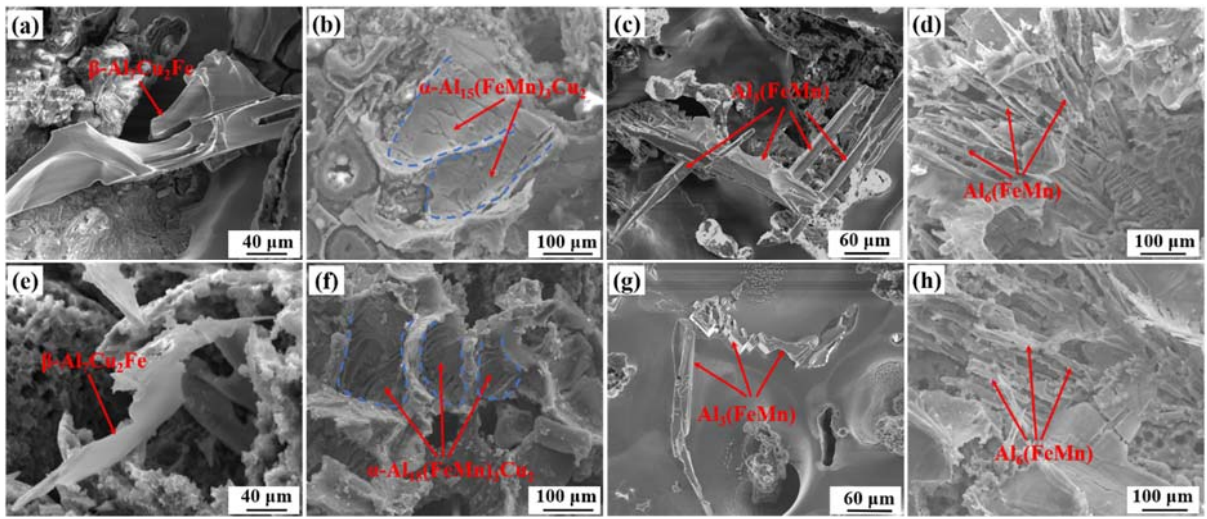

Fig. 6 The SEM images showing the deeply-etched Fe-rich phases: (a-d) without USP; (e-h) with USP; (a, e) β-$Al_7$$Cu_2$Fe; (b, f) α-$Al_{15}$(FeMn)$_3$$Cu_2$; (c, g) $Al_3$(FeMn); (d, h) $Al_6$(FeMn).

## 3.2 Effect of USP on the 3D microstructure of the alloys

In order to clearly observe the distribution of intermetallic phases in 1.0Fe alloy, the α-Al matrix was removed to show the representative 3D rendered structures of Fe-rich phases, $Al_2Cu$ and pores (Fig. 7). The complex 3D structural and large interconnectivity of network Fe-rich phases and $Al_2Cu$ were interwoven in the interdendrite of both alloys. As displayed in Fig. 7a, $Al_2Cu$-rich area and a number of large gas pores are visible in the 1.0Fe alloy without USP. Whereas, Fe-rich phases and $Al_2Cu$ are homogenous distributed in the 1.0Fe alloy with USP (Fig. 7b). Moreover, the volume fraction of pores in the 1.0Fe alloy with USP is less than those of without USP. This indicating that USP can improve the homogenous distribution capability and soundness of ingot.

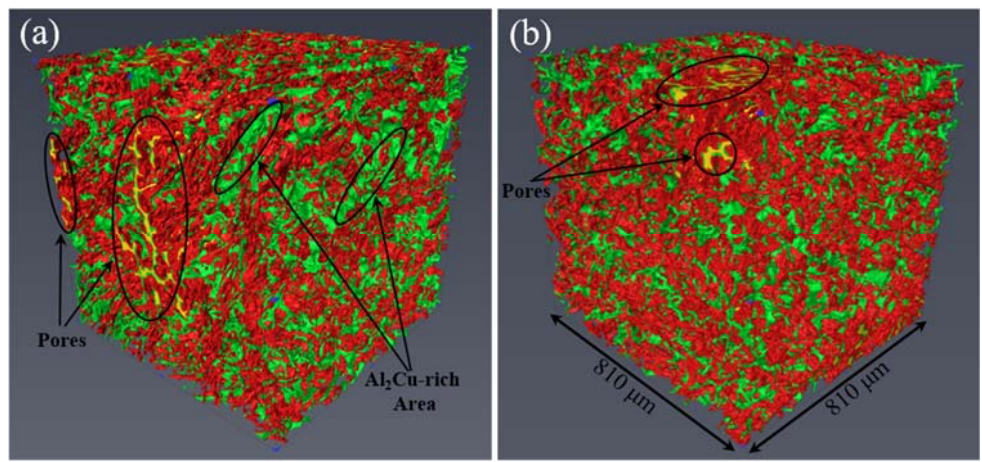

Fig. 7 Reconstructed 3D microstructures of remove the α-Al matrix in the 1.0Fe alloys without USP (a) and with USP (b).

Fig. 8 presents the reconstructed structures of Fe-rich phases obtained from SRXCT slices in the volume of $810 \times 810 \times 810$ μm³. The different colours indicated that Fe-rich phases are not fully interconnected in the present volume, which is consisting of interconnected (red colour) and dis-interconnected (other colours) particles. Fig. 8b and f indicating that the degree of interconnectivity in 0.5Fe decreased after USP. And the 1.0Fe alloy show the similar trend. The evolution of interconnectivity, equivalent diameter and specific surface area of the Fe-rich phases in the alloys without and with USP are plotted in Fig. 8. The interconnectivity of Fe-rich phases in 0.5Fe and 1.0Fe alloys without USP is 96.9% and 88.7%, while the interconnectivity of Fe-rich phases in 0.5Fe and 1.0Fe alloys with USP is 79.4% and 82.5%, respectively. The loss in the degree of interconnectivity is due to acoustic cavitation bubbles continually attack the intermetallic phases resulting in the fatigue broke and distribution dispersedly in the solidified alloys [37]. The equivalent diameter of Fe-rich phases for two alloys without USP is 137 μm and 131 μm while for with USP condition is 122 μm and 125 μm, respectively. Thus, USP process can decreased the equivalent diameter of Fe-rich phases. The evolution of specific surface area of Fe-rich phases indicates that Fe-rich phases become more compacted after USP processed. This result showing good agreement with previous observation of Al alloys [21].

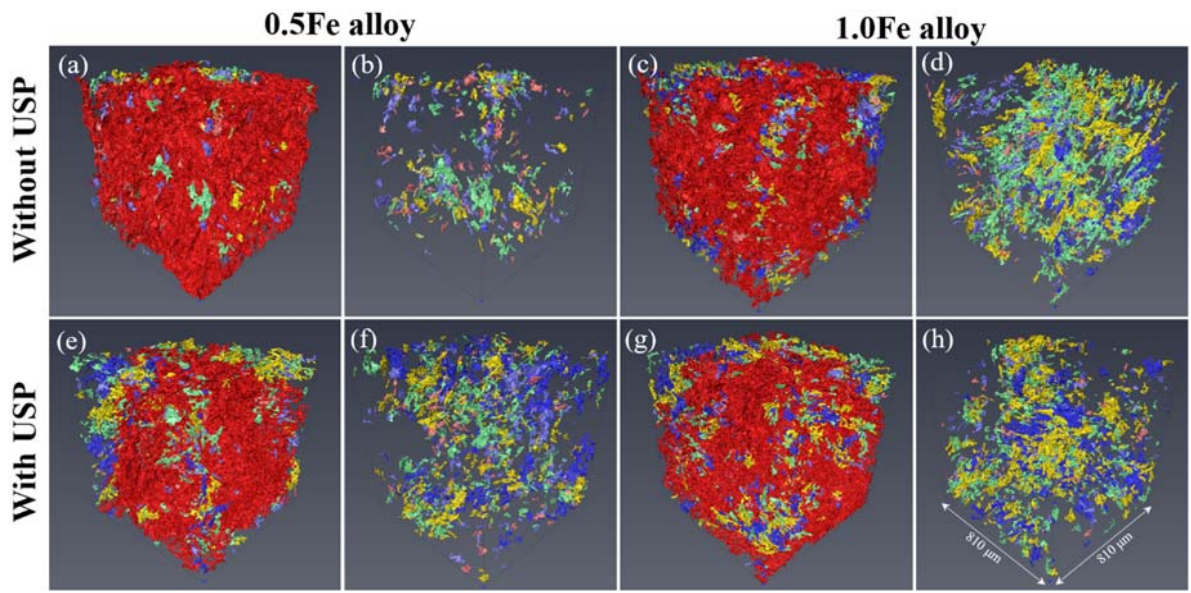

Fig. 8 3D rendered images of the Fe-rich phases in the alloys: (a-b) 0.5Fe alloy without USP; (c-d) 0.5Fe alloy with USP; (e-f) 1.0Fe alloy without USP; (g-h) 1.0Fe alloy with USP. red colour represented the interconnected particles and other colours represented the dis-interconnected particles

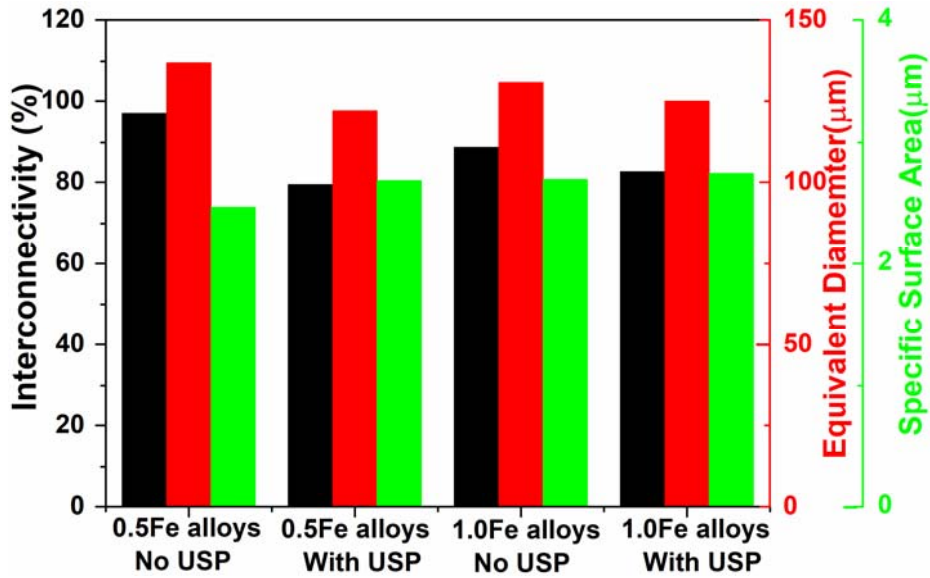

Fig. 9 Statistics of interconnectivity, equivalent diameter and specific surface area of the Fe-rich phases in the alloys with and without USP.

The volume fractions calculated for Fe-rich phases, $Al_2Cu$ and pores obtained from OM and synchrotron X-ray tomography are summarized in the Fig. 10. For the volume fraction of different intermetallic phases statistics by different methods, the value for OM is higher than those obtained from tomography. The measured area fraction obtained from OM in 2D is assumed equal to the volume fraction in 3D. Therefore, the volume fractions derived from tomography are more accurate due to the high spatial resolution of SRXCT. The data in Fig. 10 indicates that the volume fraction of Fe-rich phases and pores in the alloys without USP is higher than those of with USP. This is possible due to the effect of USP on the melting resulting in the increasing solubility of Cu in α-Al matrix [38]. For the volume fraction of Fe-rich phases, $Al_2Cu$ and pores in 0.5Fe alloy obtained from tomography, the value is 4.7%, 4.7 % and 0.9 % for the without USP condition, it decreased to 3.3 %, 4.2 % and 0.2 % for with USP condition, respectively. For both 0.5Fe and 1.0Fe alloys, the volume fraction of Fe-rich phases increased while the volume fraction of $Al_2Cu$ decreased as the Fe content increased from 0.5 % to 1.0 %. The Fe element is combined with Cu element to form Fe-rich phases through the solidification reaction in the first stage and then follow by the eutectic reaction to form Al$_2$Cu. As the Fe content increased, the Fe-rich phases consumed more Cu and less Cu left in the remainder liquid to form Al$_2$Cu [23]. Moreover, the volume fraction of pores in alloys is decreased after USP. For example, the volume fraction of pores in 1.0Fe alloys is decreased from 1.4 % without USP to 0.9 % with USP.

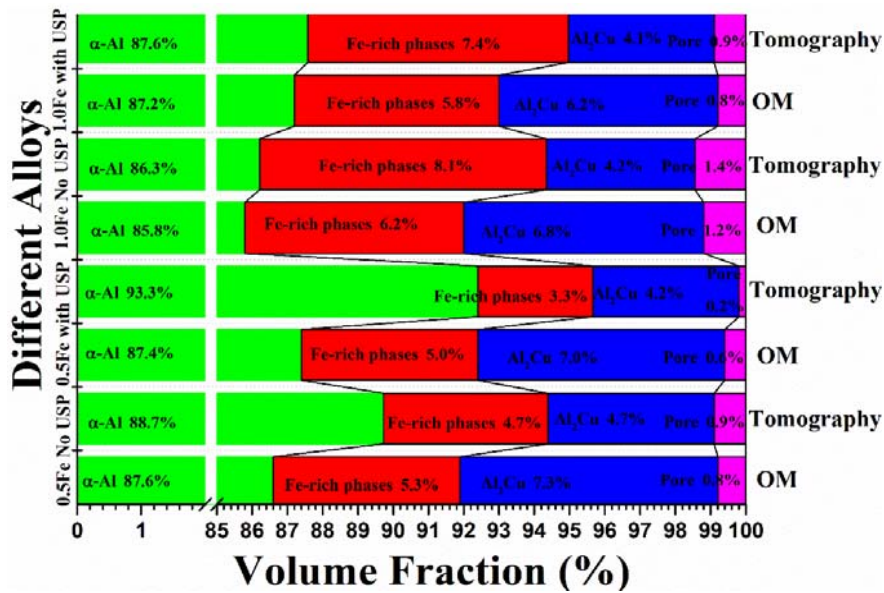

Fig. 10 Volume fractions of α-Al, Fe-rich phases, Al$_2$Cu and pore calculated from optical microscope (OM) 2D images and synchrotron X-ray tomography.

In order to study the effect of USP on the 3D morphology and size of Fe-rich phases, they were deliberately segmented to show the difference, as shown in Fig. 11. Due to the small difference in the X-ray energy spectrum absorption coefficient, it is difficult to segment different Fe-rich phases according to the similar slice contrast. Thus, the segmented Fe-rich phases (Fig.11) were carefully compared with the deep-etched SEM images (Fig. 6). As shown in Fig. 11a and b, the rod-like Al$_3$(FeMn) phase without USP exhibits several small branches, while the sample with USP sample have less branches. The size and branch of interconnected α-Al$_{15}$(FeMn)$_3$Cu$_2$ in the alloy with USP is much smaller than the alloy with USP (see Fig. 11c and d).

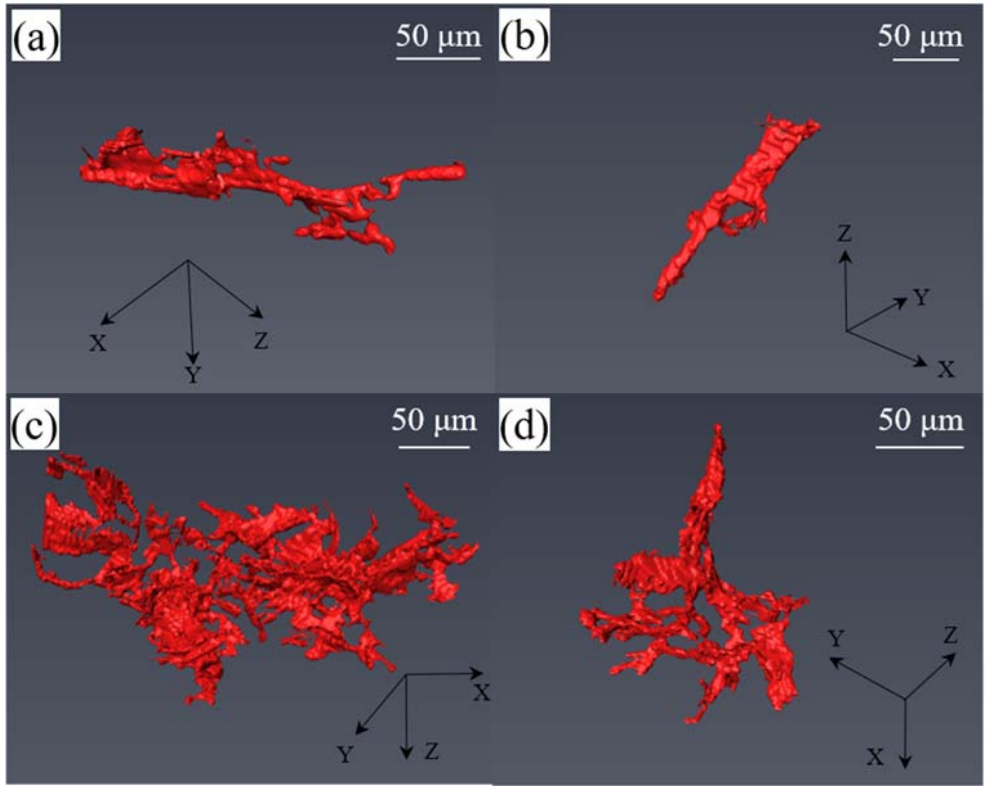

Fig. 11 Typical 3D rendered single Fe phases: (a) $Al_3(FeMn)$ without USP; (b) $Al_3(FeMn)$ with USP; (c) $\alpha$-$Al_{15}(FeMn)_3Cu_2$ without USP; (d) $\alpha$-$Al_{15}(FeMn)_3Cu_2$ with USP.

The complex 3D structure of large interconnectivity of networks $Al_2Cu$ in the 0.5Fe and 1.0Fe alloys without and with USP can be clearly seen in Fig. 11a and b. In order to distinguish the interconnected $Al_2Cu$ particles, the interconnected $Al_2Cu$ particles were painted by green colour and the dis-interconnected $Al_2Cu$ particles were painted by other colours. The interconnectivity of $Al_2Cu$ particles in 0.5Fe and 1.0Fe alloys without USP is 88.7 % and 84.6 %, it decreased to 88.7 % and 84.6 % after USP, respectively. This also indicating that USP can broke and fragmented the $Al_2Cu$ particles in the later stage solidification of eutectic reaction. In order to observe the detailed features, the enlarged typical $Al_2Cu$ particles in 0.5Fe alloy without and with USP in the volume of $129 \times 100 \times 230$ $\mu m^3$ are represented in Fig. 13. As shown in Fig. 13a and c, the high mean curvature region (red colour) is contacted with the primary $\alpha$-Al matrix. The mean length of $Al_2Cu$ particle is 25 $\mu$m for the alloy without USP, while is 14 $\mu$m for the alloy with USP (Fig. 13b and d). This is due to the USP induced the refinement of primary α-Al dendrite, Al$_2$Cu particle can only formed in the narrow room of interdendrite at the later stage of solidification. This further indicating that USP resulting in the compact 3D structure of Al$_2$Cu particle.

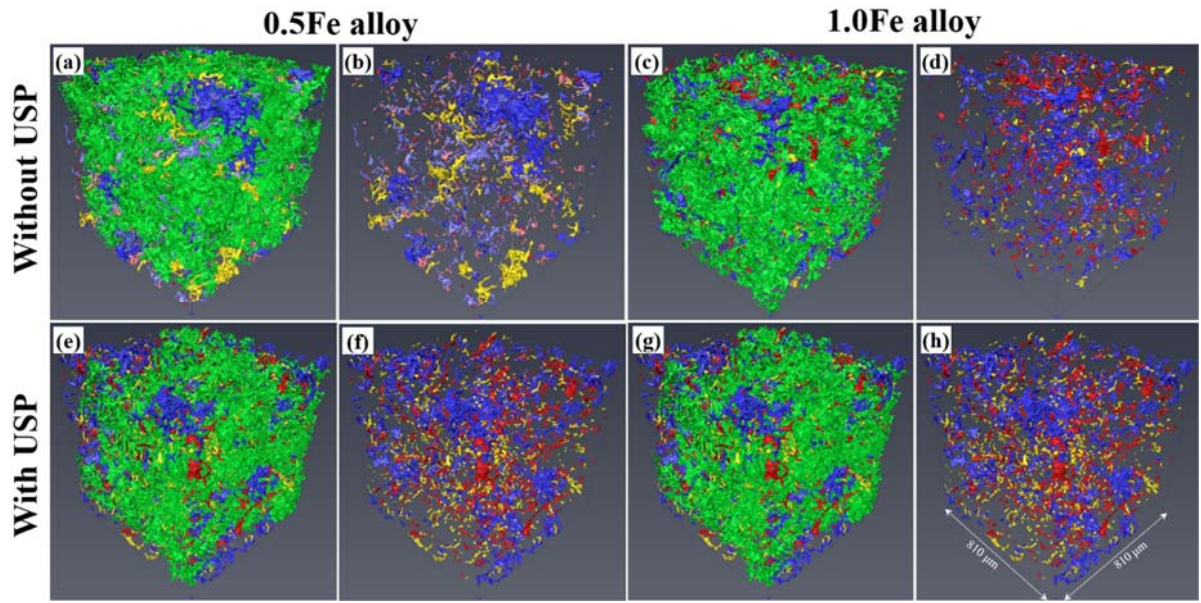

Fig. 12 Typical 3D rendered Al$_2$Cu phases: (a-b) 0.5Fe alloy without USP; (c-d) 0.5Fe alloy with USP; (e-f) 1.0Fe alloy without USP; (g-h) 1.0Fe alloy with USP. Green colour represented the interconnected Al$_2$Cu particles and other colours represented the dis-interconnected Al$_2$Cu particles.

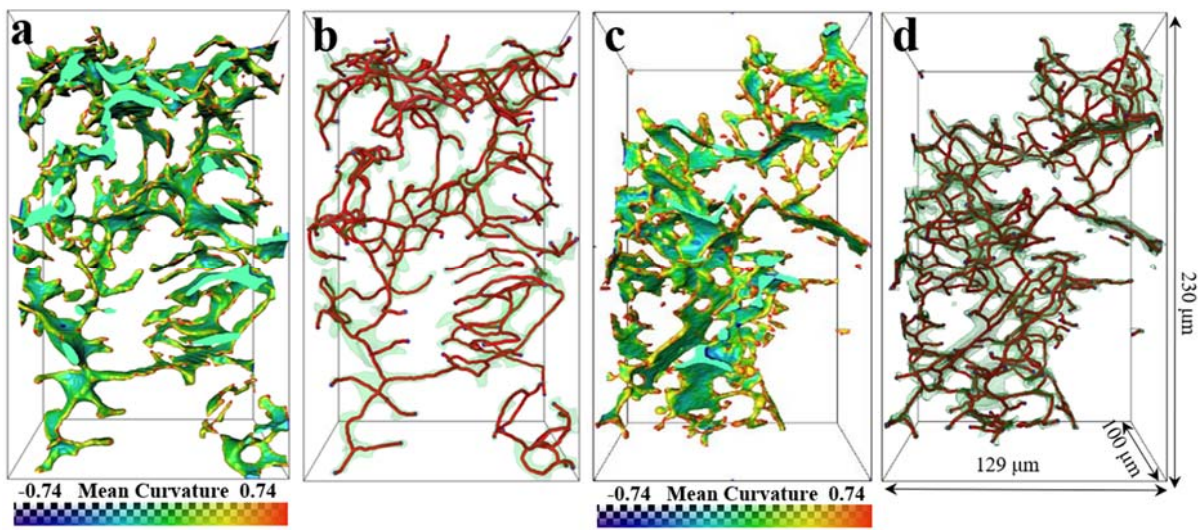

Fig. 13 The mean curvature and skeletonization of enlarged single Al$_2$Cu particles: (a, b) 0.5Fe alloy without USP; (c, d) 0.5Fe alloy without USP.

Pores are the common defects in cast Al alloys, which is usually occurs as shrinkage pores and gas pores [39]. Owing to pores is severely deteriorated the mechanical properties, including fatigue properties and ductility, of alloys. Thus, pores have been attracted by more and more researchers [15, 40]. Large interconnected shrinkage pores and globular gas pores was observed in this study, are given in Fig. 14. It can be seen that mostly pores are 3D complex shape, which were dictated by the primary α-Al matrix in the final stage of solidification [40]. Gas pores is derived from the entrapped air and release of dissolve hydrogen due to the solubility of hydrogen in Al melting is decreased with decreasing temperature [40]. Gas pores usually exhibits round shape. Fig. 14 represents the 3D reconstructed structures of pores in the alloys without and with USP. As summarized in Fig. 10, the volume fraction of pores for 0.5Fe and 1.0Fe alloys without USP are 0.9 % and 1.4 %, while for alloys with USP are 0.2 % and 0.9 %, respectively. The ultrasonic degassing results in decreasing volume fraction of porosity, this phenomenon also can be found in Refs [8-12, 41]. The enlarged 3D morphology of shrinkage and gas pores are clearly observed in Fig. 15 a-c. As shown in Fig. 15a and b, shrinkage pores exhibits the 3D complex morphology and is mainly prohibited by the primary α-Al matrix. The high mean curvature region (red colour) is contacted with the primary α-Al matrix or intermetallic phases. Their sizes vary from 20 μm to 500 μm. The 3D morphology of gas pores is round shape and with size of ~ 10 μm, as shown in Fig. 15c.

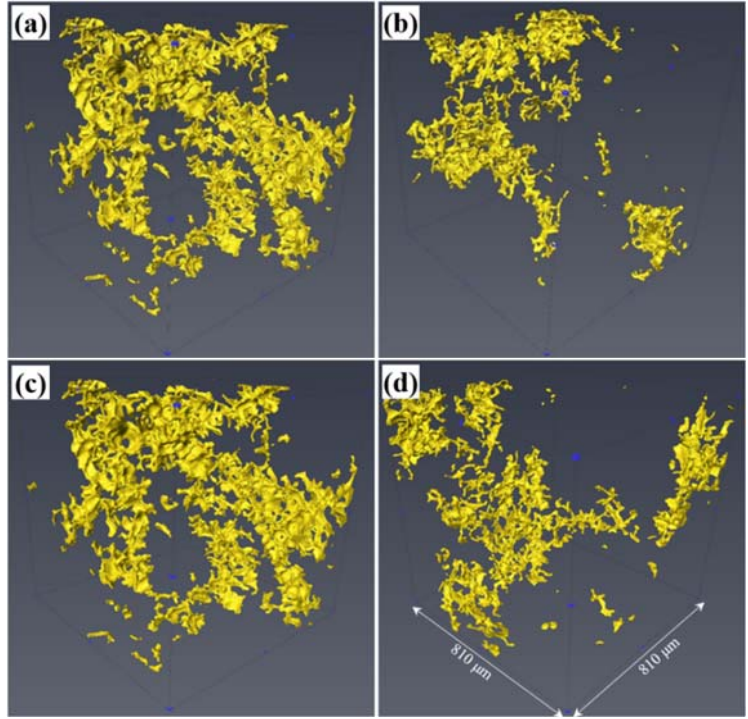

Fig. 14 Typical 3D rendered structure of pores: (a) 0.5Fe without USP; (b) 0.5Fe with USP; (c) 1.0Fe without USP; (d) 1.0Fe with USP.

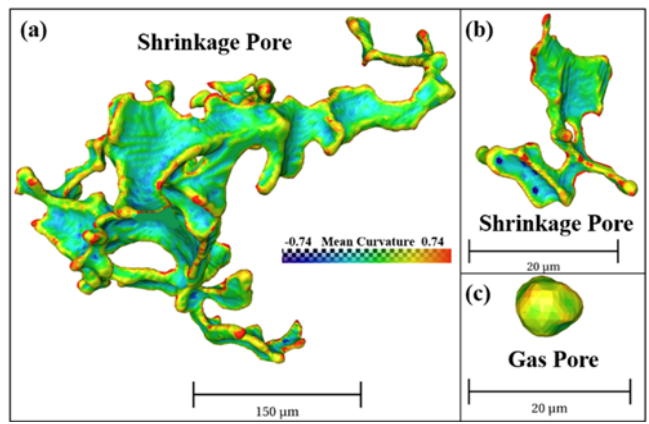

Fig. 15 The enlarged 3D morphology of pores: (a, b) shrinkage pore; (c) gas pore.

### 3.3 Refinement mechanism induced by USP

It is generally recognized [8, 9] that the following two main mechanisms resulting in the grain refinement: (1) acoustic streaming flow induced the dendrites fragmentation and bring them to form new crystals; (2) acoustic bubbles ongoing break the dendrites resulting in the fatigue fracture. A schematic diagram to illustrate the refinement mechanism induced by USP in shown in the Fig. 16.

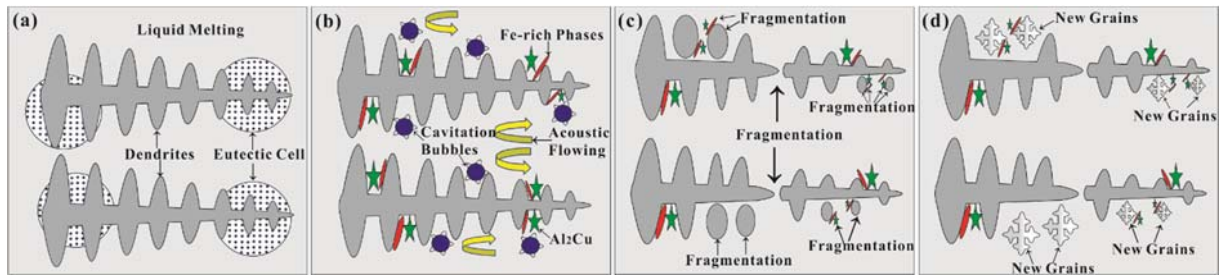

Fig. 16 Schematic diagram showing the acoustic streaming and bubbles breaking the α-Al matrix and intermetallic phases induced by USP.

### 3.3.1 Dendrites fragmentation induced by acoustic bubbles

The evidences were clearly showed in Figs. 2-5 that the refined and equiaxed grain structures can be produced after USP. It also can be concluded from Figs. 7-8 and 11-13 that the size and interconnectivity of Fe-rich and $Al_2Cu$ particles was decreased with the application of USP. The main mechanism is: i) the oscillating bubbles was cyclically attacked the primary particles at the liquid/solid interface, resulting in the fatigue fracture of primary particles [8, 9, 37]. According to the related studies reported in Refs. [37, 42], the fatigue strength of particles in melting caused by ultrasonic bubbles is about 20 ~ 30 MPa at high temperature. ii) bubble implosion occurred when the negative period of the acoustic pressure, resulting in the releasing of high-pressure shock wave to the surrounding liquid [8, 9, 37]. The imploded bubble produced many tiny bubbles or bubble fragments, which is acted as nuclei for the next cycle of bubble nucleation, expansion and implosion. The direct observation of dynamic bubble behaviour induced by USP [37, 42] provide strong evidence that the primary dendrites or secondary particles in the solid-liquid front was attacked by the ongoing and passing of chaotic and violent cavitation bubble clouds and then fragmented and detached the dendrites or particles. These fragmented dendrites and particles were immediately sweep out by the acoustic streaming flow. The sound pressure $P_A$ before the onset of cavitation can be calculated by the following equation [9]:

$$P_A = \sqrt{\frac{2W\rho c}{S}} \qquad (6)$$

where $W$ is the useful acoustic power transmitted into the melt, $\rho$ is the density of Al melting (2.43 g cm$^{-3}$ at 660 °C [28]), $c$ is the velocity of sound in Al melting (4.7 × 10$^3$ m s$^{-1}$ [29]) and $S$ is the surface area of probe (10 mm in diameter, 7.85×10$^{-3}$ m$^2$). The consumed power by the transducer is 900 W in this work. Thus, the acoustic pressure is simply estimated as ~ 1.62 MPa. Previous studies [37] showed that the cavitation threshold of Al melting was approximately 0.6-1.0 MPa (660 ~ 780 °C). The applied sound pressure in present study is higher than the cavitation threshold, will resulting in the ultrasonic cavitation bubbles in Al-Cu melt. Thus, the main refinement mechanism is the particles fragments induced by bubbles implosion and oscillating bubbles was cyclically attacking and then acoustic flow bring them to new area as nucleation sites for secondary phases (*i.e.* Fe-rich phases and Al$_2$Cu) and the primary α-Al dendrites.

### 3.3.2 Dendrites fragmentation and remelting caused by acoustic streaming flow

The main mechanism of acoustic flow in melting is remelting and deflection the dendrites and facilitate temperature and solute uniformity in the treated volume [42]. The experiment results observed by synchrotron X-ray radiography [37, 42] strongly suggests that the acoustic streaming flow not only caused the remelting of primary dendrites or secondary particles but also detached these dendrites and particles. A simple model proposed by Pilling [43] and Wang [42] to assess the primary dendrite fragmentation caused by mechanical stress of acoustic streaming flow. The equation of mechanical stress $\sigma$ is simplified as follows:

$$\sigma = \frac{6\eta V L^2}{r^3} \tag{7}$$

Here $r$ is the radius of the dendrite, $L$ is the length, $\eta$ is the dynamic viscosity, and $V$ is the flow velocity. The viscosity $\eta$ for Al-5Cu alloy at 660 °C is about $1.5 \times 10^{-3}$ Kg m$^{-1}$ s$^{-1}$, length $L$ of dendrites is equal to 2000 μm, velocity $V$ of acoustic flowing is approximately 0.5 m s$^{-1}$, radius $r$ of dendrites is assumed to 10 μm. For the dendrite radius of 10 μm, the primary dendrites mechanical breakdown needs 18 MPa stress caused by the acoustic streaming flow. If the stress is below 18 MPa, the flow only resulting in the dendrites swag like the seagrass in the ocean concurrent [42].

Acoustic enhance streaming flow is an important method on promote the temperature and solute uniformity in the Al melting [9, 42]. Usually, the thermal and temperature gradient is presented during solidification. The acoustic flow ongoing detached and fragmented the dendrites and then these fragmented particles flow with melting to new area. These particles will be acted as nuclei for the particles on the subsequent solidification. Finally, lead to the temperature and solute uniformity distribution and the microstructural refinement.

### 3.3.3 Ultrasonic degassing effect induced by acoustic streaming and bubbles

Ultrasonic degassing has been one of the most effective way of reducing hydrogen porosity. As shown in Fig. 14, the volume fraction of pores in alloys after USP is greatly decreased. The similar phenomenon was observed in the Refs. [8-12, 41]. Ultrasonic degassing, an environmentally clean and sustainable technique, uses high intensity ultrasonic vibrations to generate oscillating pressures in molten aluminum. The alternating pressure creates a large number of small cavities in the liquid. Some of these cavities grow rapidly under the influence of the alternating pressure and the unidirectional diffusion of dissolved hydrogen from the melt to the cavities. These large bubbles coagulate and float to the surface of the melt due to gravity and the acoustically induced flows in the melt [8].

## 4  Conclusions

The effect of USP on the microstructure evolution of Al-Cu-Mn-Fe alloys ere systematically investigated by conventional microscopy and synchrotron X-ray microtomography. The main results can be summarized as follows:

1). The application of USP in melting can obtained refined, equiaxed grain structures. The grain size of 0.5Fe and 1.0Fe alloys is greatly decreased from 16.9 μm, 15.8 μm without USP to 13.3 μm, 12.2 μm with USP, respectively.

2). The size, interconnectivity, equivalent diameter of network structures of secondary phases (i.e. Fe-rich phases and $Al_2Cu$) is greatly decreased after USP. This is owing to the USP induced fragmentation of primary and secondary dendrites via acoustic bubble and acoustic streaming flow. Improving the thermal and solute uniformly of alloys in the alloys.

3). USP is the effective way to reduce the volume fraction of pores in the alloys. For 0.5Fe and 1.0Fe alloys without USP are 0.9 % and 1.4 %, while for alloys with USP are 0.2 % and 0.9 %, respectively.

**Acknowledgement**

Authors gratefully acknowledge the support from Natural Science Foundation of China (51374110 and 51701075), and Team project of Natural Science Foundation of Guangdong Province (2015A030312003). We wish to thank all staff member of TOMCAT beamtime of Swiss Light Source, Paul Scherrer Institute. We also grateful to Professor Jiawei Mi's group at Hull University for providing excellent working and studying environment and High-Performance Computing facility. Financial support from the Chinese Scholarship Council (for Yuliang Zhao's PhD study at Hull University in Nov. 2016 - Nov. 2017) is also acknowledged.